\documentclass[]{raa}            

\usepackage{graphicx,times}             
\usepackage{natbib}
\usepackage{amssymb,amsmath}
\bibpunct{(}{)}{;}{a}{}{,}

\usepackage[a4paper=true,pagebackref=true]{hyperref}
\hypersetup{colorlinks = true, linkcolor = green, anchorcolor = red, citecolor = blue, filecolor = red, pagecolor = red, urlcolor = red}

\begin{document}

   \title{Contour Detection in Cassini ISS images based on Hierarchical Extreme Learning Machine and Dense Conditional Random Field}

   \volnopage{Vol.0 (20xx) No.0, 000--000}      
   \setcounter{page}{1}          

   \author{Xiqi Yang
      \inst{1,2}
   \and Qingfeng Zhang
      \inst{1,3}
   \and Zhan Li
      \inst{1,3}
   }

   \institute{Department of computer science, Jinan University, Guangzhou 510632, China; {\it tqfz@jnu.edu.cn}\\
        \and
             State Key Lab of CAD\&CG, Zhejiang University, Hangzhou 310000, China\\
        \and
        	 Sino-French Joint Laboratory for Astrometry, Dynamics and Space Science, Jinan University, Guangzhou 510632, China\\
\vs\no
   {\small Received~~20xx month day; accepted~~20xx~~month day}}

\abstract{ In Cassini ISS (Imaging Science Subsystem) images, contour detection is often performed on disk-resolved object to accurately locate their center. Thus, the contour detection is a key problem. Traditional edge detection methods, such as Canny and Roberts, often extract the contour with too much interior details and noise. Although the deep convolutional neural network has been applied successfully in many image tasks, such as classification and object detection, it needs more time and computer resources. In the paper, a contour detection algorithm based on H-ELM (Hierarchical Extreme Learning Machine) and DenseCRF (Dense Conditional Random Field) is proposed for Cassini ISS images. The experimental results show that this algorithm’s performance is better than both traditional machine learning methods such as SVM, ELM and even deep convolutional neural network. And the extracted contour is closer to the actual contour. Moreover, it can be trained and tested quickly on the general configuration of PC, so can be applied to contour detection for Cassini ISS images.
}

   \authorrunning{Xiqi Yang, Qingfeng Zhang \& Zhan Li }            
   \titlerunning{Contour Detection in Cassini ISS images based on H-ELM and DenseCRF}  

   \maketitle

%
%
\section{Introduction}           
\label{sect:intro}

Cassini ISS (Image Science Subsystem) images have been taken as an important resource for astrometry of planetary Satellites. For example,  \cite{Cooper2006Cassini,Cooper2014} derived some positions of the inner Jovian satellites from ISS images, and have done the astrometry of mid-sized Saturnian satellites from ISS images of mutual events. \cite{Tajeddine2013Astro,Tajeddine2015Cassini} reduced some observations of the main icy saturnian satellites. \cite{Zhang2018First,Zhang2019Astro} reduced some ISS images of Enceladus and Helene. Moreover, the shape and size of the Saturn and its satellites have also been obtained from ISS images, which are crucial to analyze their interior structure (\citealt{Kong201838,Kong201839,Kong201850}).

The astrometry of CCD images includes the geometry distortion correction (\citealt{Peng2017}), and matching reference stars (\citealt{Liu2018}), but they have the same prerequisite: no false detected image star. In most cases, the Saturn and its satellites are disk-resolve objects in the ISS images, which will bring some false detected image stars in the disk. Hence, detecting the contour of the disk-resolved object is important to remove false image stars. In addition, the Cassini space probe has captured more than 400,000 astronomy images, most of which require using contour detection to analyze. Thus, the speed of the contour detection algorithm is also one of the important factors considered in practical applications.

Based on different application purposes and different image conditions, researchers often use different contour detection methods, which can be divided into four categories: 

\begin{enumerate}
	\item 
	Traditional edge detection operators such as Canny, Roberts and so on. For example, \cite{Saheba2016Lunar} used improved adaptive Canny algorithm to detect lunar surface crater.  \cite{Cornet2012Edge} used an image gradient-based method to compare the contours of the Ontario Lacus to examine the displacement of the lake within five years; 
	
	\item 
	Methods driven by information theory based on well-designed features, such as the Statistical Edges method based on probability distribution (\citealt{Konishi2003Statistical}), Pb method (\citealt{Martin2004Learning}), gPb method (\citealt{Arbelaez2011}), etc.; 
	
	\item 
	Machine learning-based methods. For example, \cite{Dollar2006} proposed a supervised learning method called Boosted Edge Learning (BEL) to detect roads in satellite images. \cite{Dalal2005} used linear support vector machine (Linear-SVM) to detect the contour of pedestrian; 
	
	\item 
	Deep neural network-based methods. For example, \cite{Farabet2013} constructed a multi-scale deep convolutional neural network to implement pixel-wise classification of objects in natural images. \cite{Xie2017HED} proposed an end-to-end edge detection network called Holistically-Nested Edge Detection (HED). 
\end{enumerate}

Most of the current research focuses on the latter two categories, namely, using machine learning and deep learning techniques for contour detection. However, most of them only use natural images or other specific domain images (medical images, sensing images, etc.) as training sets. Unlike these images, Cassini ISS images have their own properties that some stars appeared as white point sources scattered in the black background and disk-resolved target has a sharply changed gray distribution to some extent. Thus, when these techniques are applied to Cassini ISS images, a large number of error detections are often caused. In addition, the training of deep neural networks often requires a lot of computing resources and time so that it may not be the best choice in practical applications. Therefore, how to accurately and quickly detect the contour of a disk-resolved object in ISS image is still a problem to be solved.

Based on our previous research (\citealt{Yang2018}), a new method based on Hierarchical Extreme Learning Machine (H-ELM) and Dense Conditional Random Field (DenseCRF) is proposed for the contour detection of Cassini ISS images. Our method combines unsupervised learning with supervised learning, using H-ELM algorithm to train a contour pixel classifier while using DenseCRF algorithm to perform back-end optimization on the contour detection results. Experiments show that our method can achieve higher accuracy and faster training speed compared with some traditional machine learning methods and even deep convolutional neural networks.


\section{Contour detection method in ISS images based on H-ELM}
\label{sect:method}
The essence of the contour detection method in ISS images based on H-ELM is to classify all the pixels in the Cassini image into two types using the best trained H-ELM model. The two types are contour pixels (denoted by 1) and non-contour pixels (denoted by 0). The overall process is divided into six parts: image preprocessing, feature selection, H-ELM network construction, network training, classification using the trained model, and back-end optimization. The key in our method is the construction and training of the H-ELM network.

\subsection{Image preprocessing}
In order to reduce the noise signals in an ISS image, preprocessing is required to perform before contour detection. In this paper, we use morphological transformation (erosion and dilation) to improve edge connections while using bilateral filtering to reduce noise and preserve edges.

\subsection{Feature selection}
Feature selection is an important step before the model training. Appropriate feature selection can make the classifier more robust. In our experiments, 34 candidate features are designed, including image gray level, Hessian feature, Kirsch operator, Robinson operator, Sobel operator, LoG operator, gradient operators, Harr-like operators and so on. Based on some common senses and some experiments, we finally select the combination of features which has the best average performance (accuracy) of the model as the feature set, namely 28 features of 3 types listed below.

\subsubsection{First-order gradients}
\begin{figure}
	\centering
	\includegraphics[width=0.50\textwidth]{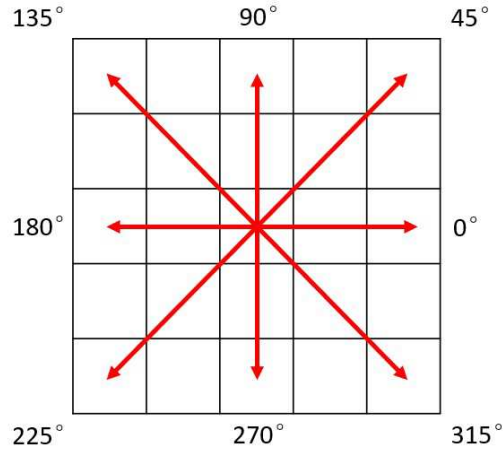}
	\caption{A pixel’s 5$\times$5 neighborhood and its eight directions used in the calculation of first-order and second-order gradient.}
	\label{direction}
\end{figure}

First-order gradient is commonly used in edge detection. In this paper, we extract the first-order gradient $g_i$ (\emph{i}=1,2,...,8) in eight directions of each pixel within its 5$\times5$ neighborhood. Figure \ref{direction} shows a pixel’s eight directions and its neighborhood. In every direction \emph{i}, we take the difference of greys between each pixel pointed by arrow and center pixel as $g_i$. Moreover, the ninth feature of a pixel is gradient amplitude $G_a$ that is calculated as follows:

\begin{equation}
G_a=\sum _{i=1}^{8} g_i \label{ga}
\end{equation}

\subsubsection{Second-order gradients}
Regarded the image as a two-dimensional discrete function, the second-order gradient is the rate of change of the gradient of the image, which can further show the edge information in the image. Unlike traditional second-order gradient, we perform the second-order derivatives in eight directions within 5$\times$5 neighborhood of one pixel (illustrated in Figure \ref{direction}). In addition, the amplitude of second-order gradient is also calculated like Equation (\ref{ga}). 

\subsubsection{Haar-like feature}
\begin{figure}
	\centering
	\includegraphics[width=0.80\textwidth]{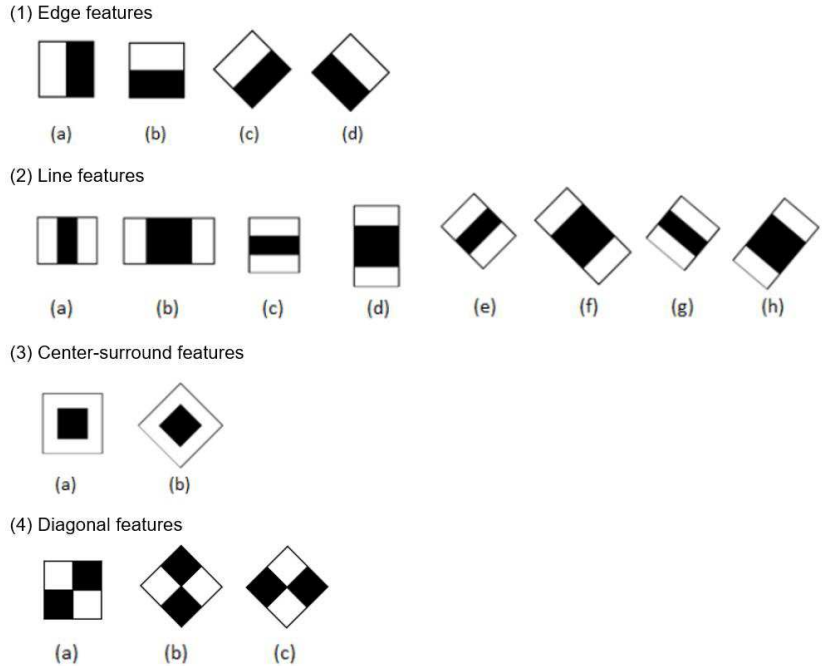}
	\caption{Haar-like features. (1) edge features. (2) line features. (3) center-surround features. (4) diagonal features. In the paper, we select line features (2a-2h) and center-surround features (3a-3b) as ten of every pixel’s features.}
	\label{haar}
\end{figure}

Haar-like feature is a simple rectangular feature similar to Haar wavelet, proposed by \cite{Viola2001,Viola2004,Lienhart2002}. There are four kinds of Haar-like features: line features, edge features, center-surround features, and diagonal features (see Figure \ref{haar}). Haar-like feature can effectively show the local grayscale change in the image, and can also be calculated fast using integral image.

In this paper, line features and center-surround features are selected. The line features (Figure \ref{haar}(2a)-(2b)) are calculated in 2$\times$3 and 2$\times$4 region, respectively. Then they are rotated at 45$^\circ$ and 90$^\circ$ to form new features (Figure \ref{haar}(2c)-(2h)). One of the center-surround features (Figure \ref{haar}(3a)) is obtained by an operator with the window size of 3$\times$3; the other (Figure \ref{haar}(3b)) is calculated in the 45$^\circ$ rotation of same region. Finally, 10 Haar-like features are extracted for each pixel.

So far, this paper has extracted 9 first-order gradient features, 9 second-order gradient features, and 10 Harr-like features. That is, we use a 28-dimensional feature vector to describe a pixel. 

\subsection{Hierarchical Extreme Learning Machine}
\subsubsection{Extreme Learning Machine}
Extreme Learning Machine (ELM) is a single-hidden-layer forward feedback network proposed by \cite{Huang2006}. In this network, the input weight and hidden layer biases are randomly generated, and the output weight is obtained by solving regularized least squares optimization. Therefore, there are only two parameters that need to be set: the number of hidden layer neurons and the activation function. Compared with the traditional machine learning algorithm and the deep neural network, it has faster training speed and more flexible parameter selection. Nowadays, ELM has demonstrated its awesome performance in many fields (\citealt{Tang2015,Mcdonnell2015,Minhas2010}).

\begin{figure}
	\centering
	\includegraphics[width=0.50\textwidth]{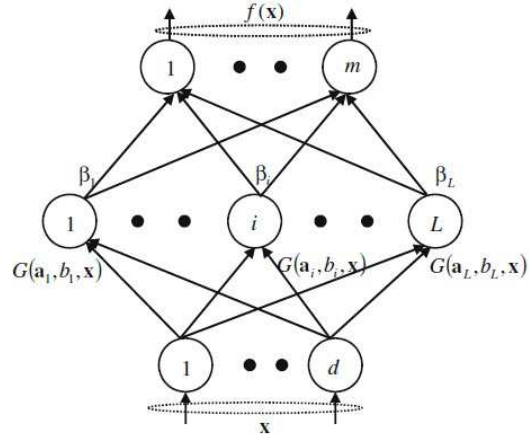}
	\caption{The structure of single-hidden-layer ELM. From top to bottom are output layer, hidden layer, input layer, respectively. \emph{m} is the output vector dimen-sion, \emph{L} is the number of hidden layer nodes, \emph{d} is the input feature vector dimension, and \emph{G} is the activation function.}
	\label{elm}
\end{figure}

The structure of ELM is shown in Figure \ref{elm} and the steps of ELM algorithm are as follows:

\begin{enumerate}
	\item Extract label matrix \emph{T} from training data
	\begin{equation}
	T=[t_1,...,t_i,...,t_N],t_i=0,1 \label{elm1}
	\end{equation}
	
	\item Randomly generate input weight $w_i$ and hidden layer bias $b_i$ of hidden layer neurons (\emph{i}=1, 2,…, N)
	
	\item Compute the output matrix \emph{H} of the hidden layer (This paper uses the sigmoid activation function)
	\begin{equation}
	H=\left[           
	\begin{array}{ccc}  
	h_1(x^{(1)}) & \cdots & h_n(x^{(1)}) \\ 
	\vdots &  & \vdots \\ 
	h_1(x^{(m)}) & \cdots & h_n(x^{(m)}) \\ 
	\end{array}
	\right]	\label{elm2}
	\end{equation}
	\begin{equation}
	\begin{aligned}
	h_j(x)&=g(\sum _{k=1}^{p}w_{ik}x_k+b_j)\\
	&=g(w_j^Tx+b_j),j\in[1,n]\label{elm3}
	\end{aligned}
	\end{equation}
	\begin{equation}
	g(z)=\frac{1}{1+e^{-z}} \label{elm4}
	\end{equation}
	
	\item According to the principle of least squares, the output weight $\beta^*$ can be calculated according to \emph{H} and \emph{T}
	\begin{equation}
	\beta^*=(H^TH)^{-1}H^TT
	\label{elm5}
	\end{equation}
	
	\item Compute the class probability of each pixel
	\begin{equation}
	y=h(x)^T\beta^* \label{elm6}
	\end{equation}
\end{enumerate}

\subsubsection{Hierarchical Extreme Learning Machine}
Although ELM has good generalization performance and approximate global approximation ability, single-hidden-layer ELM is generally used to solve simple classification problems without the feature learning and self-organization ability. Even if deep neural network has the feature learning and self-organization ability, it needs to adjust the network frequently based on the BP principle so that the training takes a long time and the parameter adjustment depends on the quality and quantity of the training samples.

Therefore, \cite{Tang2016} proposed H-ELM, which is improved on the basis of the original ELM, adding a sparse encoding layer for unsupervised feature extraction, and then using ELM for supervised classification. It not only inherits ELM's fast classification ability, but also has excellent generalization performance, thus is suitable for classification of data with multi-dimensional features, such as images.

\begin{figure}
	\centering
	\includegraphics[width=0.80\textwidth]{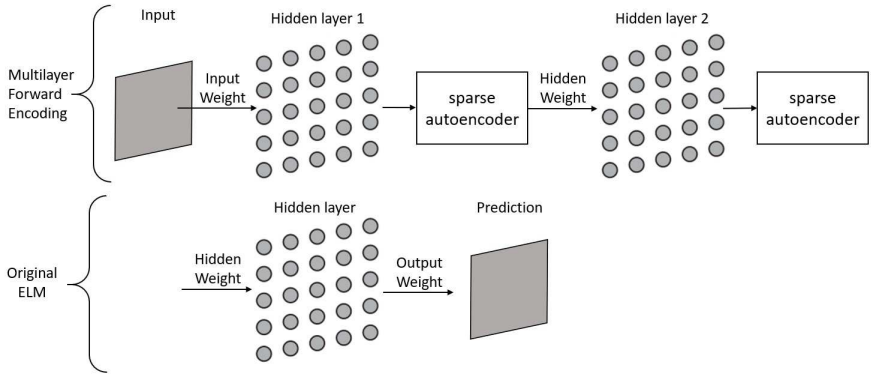}
	\caption{The network structure of H-ELM. The network includes two parts: The multilayer forward encoding and original ELM. The former will generate the learned features and the latter will take them as input data.}
	\label{helm}
\end{figure}

The H-ELM network structure used in this paper is shown in Figure \ref{helm}, which is divided into two parts: (1) Multi-layer forward encoding part, including 2 hidden layers, using the sparse autoencoder to extract the feature layer by layer, which is the unsupervised learning phase; (2) The original ELM part, including a hidden layer, classifying according to the extracted features, which is the supervised learning phase.

The sparse autoencoder makes the encoded output fit the input raw data by minimizing the reconstruction error (\citealt{Tang2016}). Thus, once the output weight of the autoencoder is obtained, it can be multiplied with the input data (i.e. feature vectors) to derive features’ compact form.

After encoding the feature vector of each pixel by using the sparse autoencoder, it will be input into the original ELM model to obtain the class probability of this pixel. 

\section{The back-end optimization method based on DenseCRF}
H-ELM can provide a set of contour pixels for each image, but the result is somewhat far from ideal. Refining the result is necessary. In this paper, we use the DenseCRF to finish the back-end optimization.

The Conditional Random Field (CRF) is a kind of conditional probability distribution model, which can convert a set of random variables into another set of random variables, and the output variables form a Markov random field (\citealt{Lafferty2002}).

Theoretically, CRF can also be used to learn the conditional distribution of each pixel’s class. However, in the contour detection problem, each pixel in the image is related to each other. For example, the pixel close to the contour pixel is more likely to be the contour pixel than the pixel far away from the contour pixel. Therefore, they will form a very dense full connected graph, whose amount of computation is large. Since DenseCRF proposed by \cite{Philipp2011} can reduce the computational complexity to sub-linearity, this paper uses DenseCRF for back-end optimization instead.

By minimizing the energy function, DenseCRF estimate the posterior distribution of the results according to the class probability predictions of the H-ELM network and the grayscale information of the image itself, thus making the contour prediction results more accurate. 

\section{Dataset preparation}
We pick out 200 ISS images to train and test our method. Every image’s size is 512$\times$512, and has one disk-resolved object. All the images show some typical features of ISS images. 130 of them are used to train our H-ELM, and 70 of them are used to test our method. The data preparation includes two steps: semi-automatic labeling and sample equalization.

\subsection{Semi-automatic labeling}
In this paper, the training and testing images are semi-automatically labeled by combining Canny operator with manual annotation. Firstly, Canny operator is used to detect all the edge pixels in the image. Then the non-contour pixels are manually removed. Finally, check whether there are contour pixels unmarked by Canny operator, if there are, label them manually.

\subsection{Sample equalization}
After sample labeling, the samples are divided into two classes: positive samples (contour pixels) and negative samples (non-contour pixels). Without further processing, the proportion of positive and negative samples is as high as 1:400, which is obviously unbalanced.

For the data imbalance problem, common solutions include under-sampling, over-sampling, using cost sensitive factors (\citealt{Barandela2004,Chawla2002SMOTE}). After experiments, the under-sampling method is adopted in this paper. By randomly removing some negative samples, the proportion of negative samples is reduced and the positive and negative sample ratio is finally close to 1:4.

\section{Experiment}
Our experiment is completed on the PC with 1.99 GHz Intel i7-8550u CPU and 8GB memory. In the training stage of the experiment, the feature vectors of each pixel in the training set are input into the H-ELM network for training. The training process is iterated for 100 times, and the optimal model parameters are saved. In the testing stage, all the pixels of the test image are input into the trained H-ELM network, and the DenseCRF is used to optimize the output result. 

\subsection{Hyper-parameter selection}
The H-ELM network has three hyper-parameters: the regular factor \emph{C}, the number of hidden layers and the number of nodes in each hidden layer. The experiment shows that the regular factor \emph{C} has a great influence on the result, so we determine the appropriate value by using the grid search method, with the search interval of [$2^{-20},2^{-18},…,2^{18},2^{20}$], and finally determine that the appropriate value is $2^{18}$. In addition, theoretically speaking, the more hidden layers H-ELM has, the stronger ability to express feature it will obtain. However, there is still no feasible method to determine the appropriate number of hidden layers and the number of hidden layer nodes. Therefore, according to our experiments and comparative analysis, this paper determines that the number of hidden layers is 3 and the number of nodes in each hidden layer is 200, 200, 1000, respectively.

\subsection{Performance metrics}
There are 4 kinds of classification results: \emph{TP} (True Positive), \emph{TN} (True Negative), \emph{FP} (False Positive), \emph{FN} (False Negative). \emph{TP} refers to the number of pixels divided into contours correctly, \emph{TN} refers to the number of pixels divided into non-contours correctly, \emph{FP} refers to the number of pixels divided into contours wrongly, and \emph{FN} refers to the number of pixels divided into non-contours wrongly.

In our experiment, we select \emph{F1-measure, Precision, Recall} as the numeric performance metrics of contour detection. \emph{F-measure} represents the weighted average of \emph{Precision} and \emph{Recall}. \emph{F1-measure} is a common form of \emph{F-measure} and becomes a standard evaluation technology in the field of contour detection and image segmentation (\citealt{Martin2004Learning,Arbelaez2011}). \emph{Precision} is the proportion of correctly classified contour pixels as the actual contour pixels, indicating the classifier's ability to correctly classify positive samples. \emph{Recall} is the recall rate of positive samples. The calculation of each performance metric is shown in Table \ref{perfmetric}.

\begin{table}
	\centering 
	\caption{Performance metrics of contour detection.}
	\begin{tabular}{cc}
		\hline
		Performance metric & Description\\
		\hline
		F1-measure & (2$\times$Recall$\times$Precision)/(Recall+Precision)\\
		Precision & TP/(TP+FP)\\
		Recall & TP/(TP+FN)\\
		\hline
	\end{tabular}
	\label{perfmetric}
\end{table}

\subsection{Experiment results}
In the training stage, the training time is about 9.004s, and the training accuracy reached 95.53\%. In the testing stage, the testing time of the single image is about 7.762s, with average \emph{F1-measure} of 58\%. The performance evaluation results of the test set are shown in Table \ref{result} (due to limited space, only some image results in the test set are listed).

\begin{table}
	\centering 
	\caption{Performance evaluation of contour detection using our method in some ISS images.}
	\begin{tabular}{cccc}
		\hline
		ISS Image No. & F1-measure & Precision & Recall\\
		\hline
		N1354204552\_1 & 0.697 & 0.535 & 1.000\\
		N1598087203\_1 & 0.683 & 0.519 & 1.000\\
		N1867018849\_1 & 0.624 & 0.530 & 0.758\\
		N1858933236\_1 & 0.614 & 0.462 & 0.914\\
		N1867018793\_1 & 0.612 & 0.489 & 0.827\\
		N1858933356\_1 & 0.608 & 0.455 & 0.915\\
		N1867018737\_1 & 0.603 & 0.500 & 0.760\\
		N1487264883\_2 & 0.603 & 0.443 & 0.945\\
		N1867018697\_1 & 0.588 & 0.461 & 0.812\\
		\textbf{Average} & \textbf{0.580} & \textbf{0.656} & \textbf{0.537}\\
		\hline
	\end{tabular}
	\label{result}
\end{table}

\subsubsection{Comparison with traditional edge detection operators}
\begin{figure}
	\centering
	\includegraphics[width=0.80\textwidth]{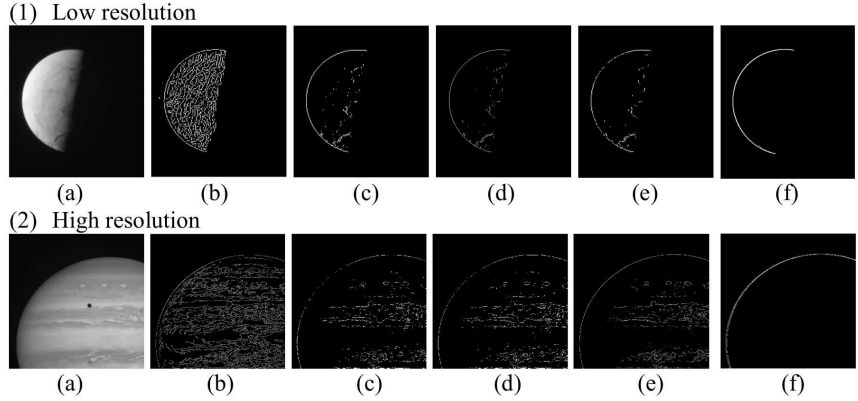}
	\caption{Contour detection results comparison in Cassini ISS images by using different methods. (a) Original Image. (b) Canny. (c) Sobel. (d) Roberts. (e) Prewitt. (f) Our method. }
	\label{traditional}
\end{figure}

Figure \ref{traditional} shows the results of different edge detection operators on the observed object with different resolutions. As we can see, Canny, Roberts and Prewitt operator (corresponding to Figure \ref{traditional}(b)-(e) respectively) bring lots of false detection, and the edge connectivity of the observed object is unsatisfactory. However, our method (Figure \ref{traditional}(f)) can extract the contour of the observed object well, while keeping the better edge connectivity.

\subsubsection{Comparison with Support Vector Machine (SVM)}
Support Vector Machine (SVM) is a Machine learning algorithm proposed by \cite{Vapnik1998} on the basis of statistical learning theory, indicating significant advantages in many pattern recognition problems like small sample problem, and achieving good results in many fields such as handwritten recognition, biological information and so on. Therefore, this paper takes SVM algorithm as the representative of statistical learning algorithm and performs comparative experiments on our dataset.

\begin{figure}
	\centering
	\includegraphics[width=0.80\textwidth]{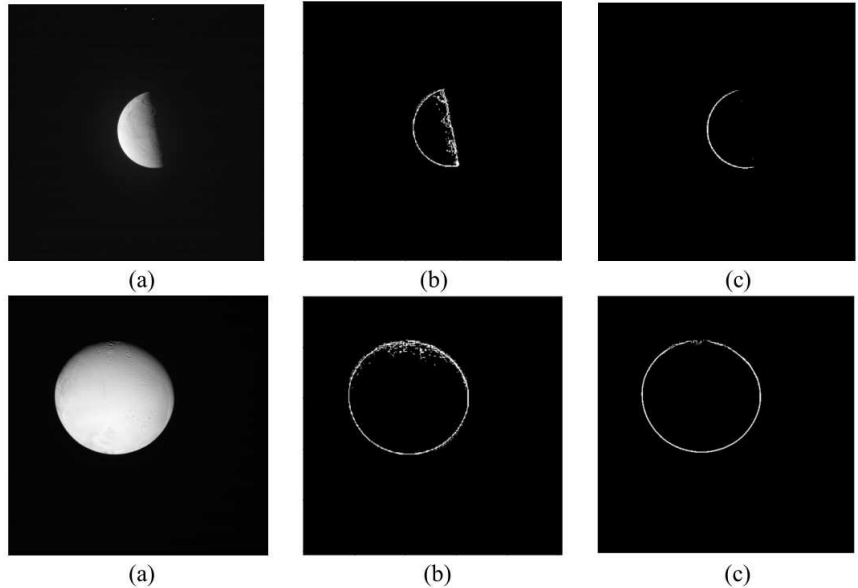}
	\caption{Contour detection results comparison in Cassini ISS images by using different methods. (a) Original Image. (b) SVM. (c) Our method. }
	\label{svm}
\end{figure}

Table \ref{perfcomp} show the difference of performance between the SVM algorithm and our method. Obviously, the training time of our method is shorter than SVM and our method is superior to SVM in all performance metrics. As is shown in Figure \ref{svm}, although SVM can extract fine contour in some way, it is not as good as our method in some inner details processing. It should note that the top (c) in Figure \ref{svm} shows our method don’t detect the terminator while SVM finds it. It is because we take the terminator pixels as non-contour ones when we label images. In fact, we use contour pixels to determine the center of disk-resolved object in our astrometry of ISS images, the terminator pixel is useless. 

\begin{table}
	\centering
	\caption{Performance comparisons among SVM, ELM, CNN and our method.}
	\begin{tabular}{ccccc}
		\hline
		& Training time (s) & F1-measure & Precision & Recall\\
		\hline
		ELM & 7.573 & 0.30 & 0.18 & 0.89\\
		SVM & 112.270 & 0.21 & 0.12 & 0.81\\
		CNN & 6138.948 & 0.12 &	0.07 & 0.39\\
		Our method & \textbf{9.004} & \textbf{0.58} & \textbf{0.66} & \textbf{0.54}\\
		\hline
	\end{tabular}
	\label{perfcomp}
\end{table}

\subsubsection{Comparison with original ELM}
In order to prove the superiority of unsupervised learning stage in H-ELM, we perform a comparative experiment between H-ELM and original ELM. The number of hidden layer nodes in original ELM is equal with the last layer (namely ELM classification layer) in H-ELM, both of which are 1000.

\begin{figure}
	\centering
	\includegraphics[width=0.80\textwidth]{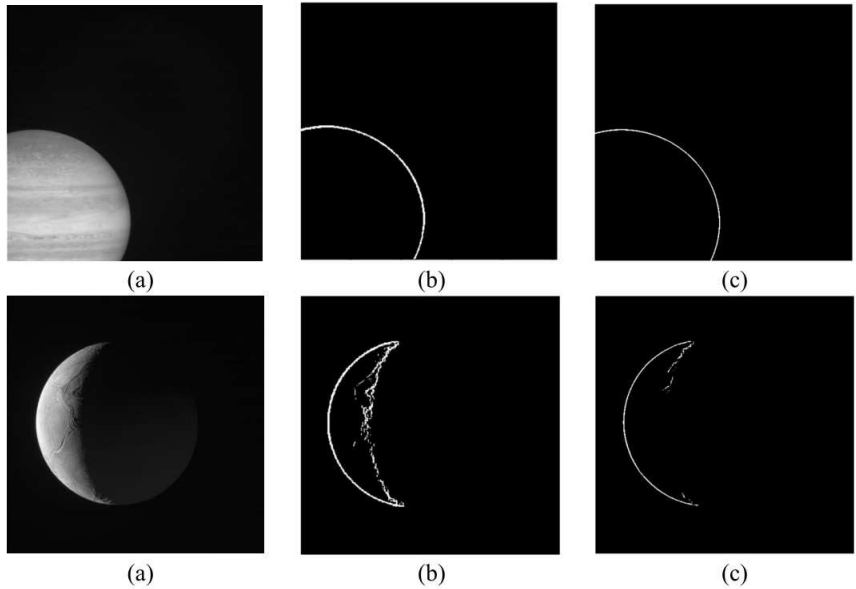}
	\caption{Contour detection results comparison in Cassini ISS images using different methods. (a) Original Image. (b) Original ELM. (c) Our method.}
	\label{elmcomp}
\end{figure}

As can be seen from Table \ref{perfcomp}, original ELM algorithm has the shortest training time and is slightly better than SVM algorithm in all performance metrics, which shows the superiority of ELM algorithm in some way. Of course, it can also be seen that both ELM and SVM are not as effective as our method. Figure \ref{elmcomp} shows the comparison results of original ELM and H-ELM on different observed objects. It can be found that, no matter in the simple or complex surface condition of the observed object, the method proposed in this paper is superior to original ELM in the precision of the extracted contour.

\subsubsection{Comparison with Convolution Neural Network (CNN)}
\begin{figure}
	\centering
	\includegraphics[width=0.50\textwidth]{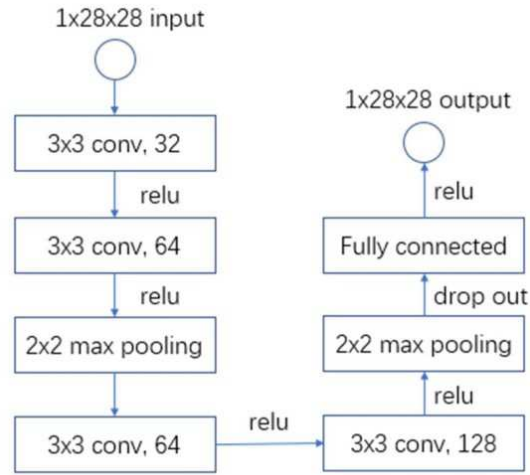}
	\caption{The CNN architecture in the comparison experiment.}
	\label{cnnarch}
\end{figure}

In order to compare our method with the currently popular deep learning method, we use the most common neural network in deep learning, namely Convolutional Neural Network (CNN) for comparative experiment. The adopted CNN architecture includes four 3$\times$3 convolution layers, two 2$\times$2 max pooling layers and a full connected layer. The activation function is relu, and the input and output are both 28$\times$28 image blocks, as shown in Figure \ref{cnnarch}.

\begin{figure}
	\centering
	\includegraphics[width=0.80\textwidth]{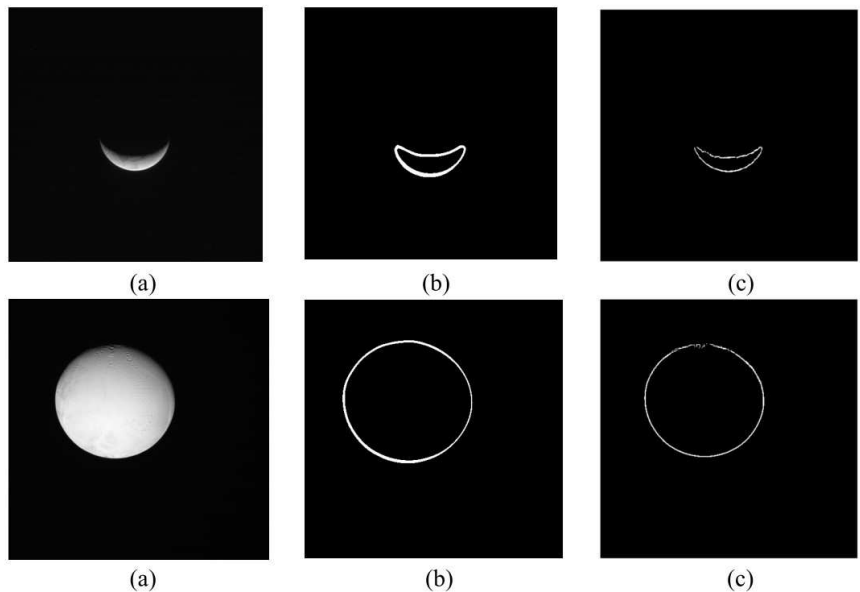}
	\caption{Contour detection results comparison in Cassini ISS images by using different methods. (a) Original Image. (b) CNN. (c) Our method. }
	\label{cnncomp}
\end{figure}

As is shown in Table \ref{perfcomp}, compared with other machine learning algorithms, the training of CNN requires a lot of time, which is also the drawback of all deep learning methods. Although CNN can extract a relatively complete contour (see Figure \ref{cnncomp}), it is far inferior to our method both in average performance metric and the precision of extracted contour. In other words, CNN can be used for rough contour detection, but it may not be the best choice for work requiring high precision of contour (such as calculating the center of celestial body according to its contour).

\subsubsection{Comparison with H-ELM (without optimization)}
\begin{figure}
	\centering
	\includegraphics[width=.8\textwidth]{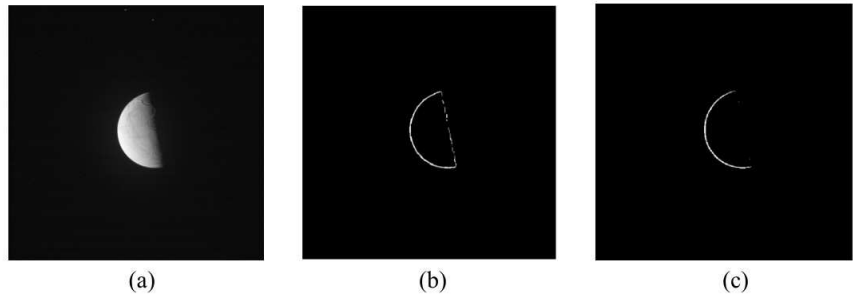}
	\caption{Contour detection results comparison in Cassini ISS images by using different methods. (a) Original Image. (b) H-ELM (without optimization). (c) Our method. }
	\label{helmcomp}
\end{figure}

Figure \ref{helmcomp} shows the comparison between the H-ELM algorithm (without optimization) and our method. Obviously, after the optimization using DenseCRF, the inner contour (terminator) of the observed object can be effectively removed, so that only the outer arc-like contour can be retained.

\section{Conclusion}
A new method for contour detection of Cassini ISS images based on H-ELM and DenseCRF is proposed in this paper. On the one hand, our method inherits the advantages of H-ELM, obtaining feature learning and self-organization ability of deep neural network while keeping the fast training ability of ELM. On the other hand, we take the implicit relation between the classes of pixels in the contour image into consider, using DenseCRF algorithm to optimize the contour results.

Through experiments, we proved that the proposed method has the following advantages:
\begin{itemize}	
	\item Extremely short training time
	\item Strong generalization ability
	\item The back-end optimization part can effectively remove part of the inner contour and retain the outermost contour of the celestial body.
\end{itemize} 

In a word, the method proposed in this paper is available for the contour detection of disk-resolved object in Cassini ISS images.

However, it should be mentioned that, at present, there are many machine learning (ML) methods. To find a better ML method than the one in the paper is our further study in the future. Compared with the traditional methods, ML methods can eliminate false contour detection to a great extent, and give a more accurate contour. It is because ML methods can learn the pattern hidden in the image, thus making solution close to manual labeled result as much as possible. Obviously, the contour extracted using ML methods will have almost no noise, and be closer to the actual contour of the object. Therefore, we believe that ML will be a promising way to solve the problem of contour detection.


\begin{acknowledgements}
This work was partly supported by National Natural Science Foundation of China (Grant No. U1431227, 11873026), Natural Science Foundation of Guangdong Province, China (Grant No. 2016A030313092) and the Fundamental Research Funds for the Central Universities (Grant No. 21619413).
\end{acknowledgements}

\bibliographystyle{raa}
\bibliography{bibtex}

\end{document}